\documentclass[preprint,nofootinbib]{revtex4}
\usepackage{epsfig}
\usepackage{amssymb}

\begin{document}
\begin{titlepage}

\begin{center}
{\Large \bf Measuring the effects of Loop Quantum Cosmology in the CMB data\footnote{Essay selected with honorable mention from Gravity Research Foun
dation (2017) - Awards Essays on
Gravitation\\$^{2}$svasil@academyofathens.gr\\$^{3}$vkamali@basu.ac.ir\\$^{4}$Mehrabi@basu.ac.ir}}


\vskip 0.5cm {\bf Spyros  Basilakos$^{a,2}$, Vahid Kamali$^{b,3}$, Ahmad Mehrabi$^{b,4}$}
\end{center}

{\small
\begin{quote}
\begin{center}
$^a$Academy of Athens, Research Center for Astronomy and Applied
Mathematics, Soranou Efesiou 4, 11527, Athens, Greece\\
$^b$Department of Physics, Bu-Ali Sina University, Hamedan 65178,
016016, Iran
\end{center}
\end{quote}
\vspace{0.2cm}
\centerline{(Submission date: March 30, 2017)}
}
\vspace{0.2cm}
\centerline{\bf Abstract}
\bigskip
In this Essay 
we investigate the observational signatures of Loop Quantum Cosmology (LQC) 
in the CMB data.
First, we concentrate on the dynamics of LQC 
and we provide the basic cosmological functions.
We then obtain 
the power spectrum of scalar and tensor perturbations
in order to study the performance of LQC against the latest CMB data. 
We find that LQC provides a robust prediction 
for the 
main slow-roll parameters, 
like the scalar spectral index and the tensor-to-scalar fluctuation ratio, 
which are in excellent agreement 
within $1\sigma$ with the values recently measured by the 
Planck collaboration. 
This result indicates that LQC can be seen as an alternative 
scenario with respect to that of standard inflation.




\vspace{0.3cm}

\end{titlepage}

\pagestyle{plain} \baselineskip 0.75cm
Recent studies of the Cosmic Microwave Background
(CMB) 
have opened 
a new window for early cosmology. Specifically, 
based on Planck data \cite{Ade:2015lrj,Keck15} it has been 
found that the inflationary models which are in agreement with the data 
are those with very low tensor-to-scalar fluctuation ratio $
r = \frac{P_t}{P_s} \ll 1$, with a scalar spectral index $n_s \simeq 0.96 $
and no appreciable running. Actually, the upper bound imposed by Planck 
team~\cite{Ade:2015lrj}, on this ratio, as a result of the
non-observation of B-modes, is $r < 0.11$ which implies that 
 $H\lesssim 10^{-5}M_p$, where $M_{p} \simeq 1.22 \times 10^{19}$Gev. 

After a long period 
of the successful inflationary paradigm, cosmology still lacks a 
framework in which 
the universe at Planck scale  
smoothly evolves to the CMB era.
In the light of the
Planck results\,\cite{Ade:2015lrj}, a heated debate 
is taking place in the literature
about the implementation of LQC to CMB data. 
Recently, Ashtekar and Gupt \cite{Gupt17} 
using various correlation functions for scalar perturbations 
found that LQC is favored by Planck, while  
standard (cold) inflation
can not accommodate the data at large angular scales ($l\le 30$).
The intense discussion is going on and 
the aim of the present Essay 
is to contribute to this debate.
Here we focus on 
the dynamical behavior of the effective Loop Quantum Gravity (LQG) theory 
via the Hubble expansion, and investigate the performance of LQC against the
latest Planck data in the slow-roll regime.

\vspace{0.5cm}
{\it Loop Quantum Cosmology}:
It is the framework that implements the basic 
cosmological principles in LQG theory
\cite{Boj2005} in which canonical quantization of gravity is given in 
terms of the so called Ashtekar-Barbero
connection variables \cite{Ashtekar:2003hd}. 
Specifically, the phase space of
classical general relativity can be spanned by conjugate
variables $A_j^i$ (connection) and $E_i^j$ (triad) on a
$3-manifold$ $\mathcal{M}$ which encodes curvature and spatial
geometry respectively. 
At the level of LQC
due to the homogeneous and isotropic symmetries the phase space is
characterized by a single
connection $c$ and a single triad $p$, while 
the Poisson bracket is given by 
$\{c,p\}=\frac{8\pi\gamma}{3M_p^2}$,
where $\gamma \simeq 0.2375$ 
is the dimensionless Barbero-Immirzi parameter.
In the context of FRW metric the above variables take the form
$c=\gamma\dot{a}$ and $p=a^2$, where $a(t)$ is the scale factor of the universe.
Based on the pair $\{c,p\}$ 
Ref.\cite{Singh:2005xg} proposed  
an effective theory of LQG which is appropriate for cosmology. 
In particular, the effective Hamiltonian is given by 
$\mathcal{H}_{\rm eff}=-\frac{3\sqrt{p}}
{\gamma\overline{\mu}^2}\sin^2(\overline{\mu}c)+\mathcal{H}_m$, where
$\mathcal{H}_m$ is the matter Hamiltonian and 
$\overline{\mu}$ 
is associated to the minimal area of LQG. Using the 
effective Hamilton equation 
$\dot{p}=\{p,\mathcal{H}_{\rm eff}\}=-\frac{\gamma}{3}\frac{\partial
\mathcal{H}_{\rm eff}}{\partial c}$
together with the 
Hamiltonian constraint ($\mathcal{H}_{\rm eff}\equiv 0)$ we 
obtain the equations of motion from which we define the 
Friedmann equation, 
\begin{equation}
 \label{newfried}
 H^2=\frac{\kappa}{3}\,\rho\,\left(1-\frac{\rho}{\rho_{c}}\right),
\end{equation}
where $H=\frac{\dot a}{a}$,
$\rho$ is the total energy density, 
$\rho_{c}=\sqrt{3}\,/(16\pi^2 G^{2}\gamma^3)$ is the 
critical loop quantum density and 
$\kappa=8\pi G = {8 \pi}{M^{-2}_{p}}$.
Evidently, the Hubble parameter in LQC modifies the standard form of 
Friedmann equation.
Bellow, we present the compatibility of 
the early phase of LQC theory, via slow roll regime, 
with warm inflationary scenario.


\vspace{0.5cm}
{\it Slow-roll regime:}
In fact we can regard 
the primeval inflationary phase of 
the early universe within a LQC framework 
as the effective action of a tachyon 
field $\phi$ in the early times \cite{Li:2008tg}.
Therefore, the total density is written as $\rho=\rho_{\phi}+\rho_{\gamma}$ in which
$\rho_{\phi}$ and $\rho_{\gamma}$
are the corresponding tachyon field and radiation densities.
Armed with the scalar field language we use the 
action of Gibbons \cite{Gibbons:2002md} 
and the energy momentum tensor 
$T^{\mu}_{\nu}={\rm diag}(-\rho,p,p,p)$ in order to
derive the conservation law of the total density 
${\dot \rho}+ 3 H (\rho+p)=0$, which reduces to the following set of equations
\begin{equation}\label{ss3}
\dot{\rho_{\phi}}+3\,H\,(\rho_{\phi}+p_{\phi})=-\Gamma\dot{\phi}^{2},\;\;\;\;\;\;
\dot{\rho}_{\gamma}+3H(\rho_{\gamma}+p_{\gamma})=\Gamma\dot{\phi}^{2},
\end{equation}
where $p=p_{\phi}+p_{\gamma}$.
The tachyon field decays to radiation \cite{del2009} and  
$\Gamma$ is the dissipation factor in unit of $M^{5}_{pl}$. 
Regarding the quantities of the fluid components appear
in Eq.(\ref{ss3}) we have $p_{\gamma}=\frac{\rho_{\gamma}}{3}$,
$\rho_{\phi}=\frac{V(\phi)}{\sqrt{1-\dot{\phi}^2}}$, $p_{\phi}=-V(\phi)\sqrt{1-\dot{\phi}^2}$
and $V(\phi)$ is the effective potential.
Clearly, the current scenario resembles the conditions of warm inflation. 
Unlike standard inflation, here 
the scalar field
is allowed to interact with other light fields, implying that radiation 
production occurs during the slow-roll
period and hence reheating is avoided \cite{Ber}.
We remind the reader that in the case of warm inflation
the condition $T > H$ 
is satisfied, 
which means that the origin of density perturbations
is thermal instead of quantum. Prior to the slow roll era 
the energy density of the 
scalar field
dominates
the cosmic fluid ($\rho_{\phi}\gg \rho_{\gamma}$), hence
the modified 
Friedmann equation 
Eq.(\ref{newfried}) is well approximated by
\begin{equation}\label{sp5}
 H^{2} \simeq \frac{\kappa}{3} \rho_{\phi} \left(1  -\frac{\rho_{\phi}}{\rho_{c}}\right).
 \end{equation}
Within this framework, using 
Eq.(\ref{ss3}) and Eq.(\ref{sp5}) we arrive at 
\begin{equation}
\label{dotphi.1}
 \dot \phi^{2} \simeq - \frac{2\dot H }{ \kappa  \rho_{\phi} (1+R)} \left(1 - \frac{12 H^{2}}{\kappa \rho_{c}}\right)^{-\frac{1}{2}},
\end{equation}
where $R\equiv \Gamma/3H\rho_{\phi}$.
Now, if we consider that the quantity 
$\Gamma \dot \phi^{2}$ varies adiabatically then 
the radiation component evolves slowly 
($\dot{\rho }_{\gamma}\ll\Gamma\dot{\phi}^{2}$)
and thus combining 
Eq.(\ref{ss3}),
Eq.(\ref{sp5}) and Eq.(\ref{dotphi.1}) we find
\begin{equation}
\label{sp55}
\rho_{\gamma} \simeq \frac{\Gamma\dot{\phi}^{2}}{4H}
\simeq \frac{- \Gamma \dot{H}}{2 \kappa
H(1+R)\rho_{\phi}} \left(1 - \frac{12 H^{2}}{\kappa \rho_{c}}\right)^{-\frac{1}{2}}.
\end{equation} 
Assuming that the tachyon field decays adiabatically, that is, in such a 
way that the specific entropy of the massless particles remains constant,
the corresponding temperature is given by Kolb and Turner \cite{Kolb} 
via $T=\left(\rho_{\gamma}/C_{\gamma}\right)^{1/4}$. 
Notice, however that the temperature does not satisfy the standard scaling law,
$T \propto a^{-1}$. Indeed, in the present scenario 
with the aid of Eq.(\ref{sp55}) 
the evolution law for $T$ becomes
\begin{equation}\label{tem}
T \simeq \left[  \frac{-\Gamma\dot{H}}{2 \kappa  C_{\gamma} H(1+R)\rho_{\phi}} \right]^{\frac{1}{4}}\left(1 - \frac{12 H^{2}}{\kappa \rho_{c}}\right)^{-\frac{1}{8}},
\end{equation}
where $C_\gamma=\frac{\pi^2 g_*}{30} \simeq 70$. 
Solving Eq.({\ref{sp5}) 
for the tachyon density and using at the same time 
Eq.(\ref{dotphi.1}) and $\rho_{\phi}=\frac{V(\phi)}{\sqrt{1-\dot{\phi}^2}}$
we obtain 
\begin{equation}\label{pot}
V(\phi)\simeq \frac{\rho_{c}}{2}
\left({1- \sqrt{1 - \frac{12H^{2}(\phi)}{\kappa \rho_{c}}}}\right) 
\times {\left[1+\frac{2\dot H(\phi) }
{ \kappa  \rho_{\phi} (1+R)} \left(1 - \frac{12 H^{2}(\phi)}
{\kappa \rho_{c}}\right)^{-\frac{1}{2}} \right]^{\frac12}}.
\end{equation}
In the slow roll regime $\rho_{\phi} \approx V(\phi)$ 
(or ${\dot \phi}^{2}\ll 1$) the above equation 
reduces to $V(\phi)\simeq \frac{\rho_{c}}{2}
\left({1- \sqrt{1 - \frac{12H^{2}}{\kappa \rho_{c}}}}\right)$.

At this point we are ready to provide the spectral indices
$n_{s}\equiv 1+\frac{d{\rm ln}P_{s}}{d{\rm ln}k}$ 
with $d{\rm ln}k=-dN$ 
and $r\equiv \frac{P_{t}}{P_{s}}$.
Here the capital $N$ denotes the number of e-folds 
$N=\int_{t}^{t_{end}} Hdt$, where $t_{end}$ is the time at the end of
inflation.
The pair $(P_{s},P_{t})$
is defined in terms of the aforementioned cosmological quantities.
Indeed, the amplitude of tensor fluctuations is given by 
$P_{t} = 8 \kappa \left(\frac{H}{2 \pi}\right)^{2}$ 
and the power spectrum of scalar fluctuations 
is written as $P_{s} = \frac{H^{2}}{\dot \phi^{2}} \delta \phi^{2}$. 
Since the nature of scalar perturbations is thermal we utilize 
$\delta\phi^2\simeq\frac{k_F T}{2 M_{pl}^4\pi^2}$ \cite{Hall:2003},
which is valid in the high dissipation regime ($R\gg 1$). 
The wave number 
$k_F=\sqrt{\frac{\Gamma H}{V}}=H\sqrt{\frac{\Gamma}{3HV}}\geq H$ 
provides the freeze-out scale at which 
the dissipation damps out to thermally excited 
fluctuations of scalar field  
($\frac{V''}{V'}<\frac{\Gamma H}{V}$) \cite{Taylor:2000ze}. 
Inserting the wave number in $P_{s}$
we find $P_{s} = \frac{\kappa^2 H^{5/2} \Gamma^{1/2} T}{128 \pi^{4}  V^{1/2} \dot \phi^{2}}$.

In order to proceed 
with the analysis we need to know the functional form of $\Gamma$.
In fact, knowing $\Gamma$ 
we can solve the system of equations (\ref{sp5})-(\ref{pot})
which means that the main 
parameters $(P_{s},P_{t})$ can be 
readily calculated, and from them $(n_s,r)$ immediately ensue. 
Here 
we treat the dissipation parameter 
as follows $\Gamma=C_{\phi}T^{m}\phi^{1-m}$ \cite{Bas11}, where 
$C_{\phi}$ is constant.
Depending on the values $m$ we have: 
(a) for $m=-1$ we get $\Gamma=C_{\phi}T^{-1}\phi^{2}$ which corresponds 
to the 
non-SUSY model \cite{Berera:1998gx,Yokoyama:1998ju},
(b) $\Gamma = C_{\phi} \phi$ ($m=0$) corresponds to SUSY \cite{Berera:1998gx}
and (c) for $m=1$ we obtain 
$\Gamma = C_{\phi} T$ \cite{Berera:2008ar}.

%
%

\begin{figure*}
\centerline{\epsfig{file=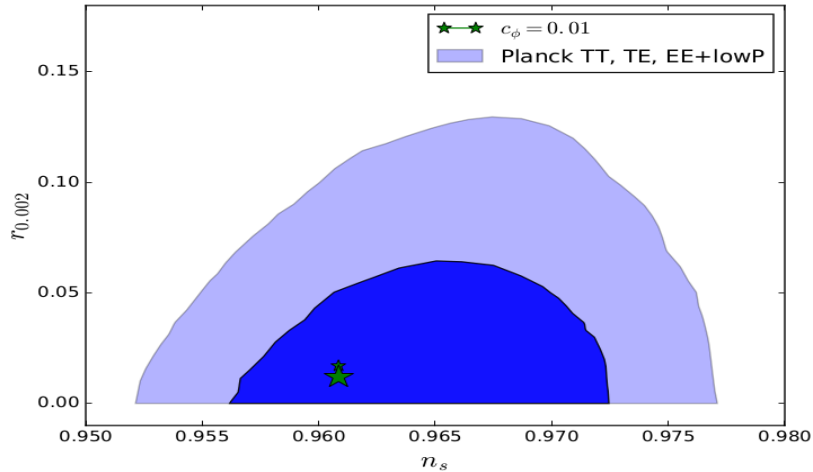,width=120mm, height=70mm} \hskip 0.1in}
\caption{The $(n_{s},r)$ diagram for LQC with $\Gamma \propto T$ using 
$N=60$ (big star) and $N=50$ (small star).
The $1 \sigma$ and $2 \sigma$ contours borrowed from 
Planck \cite{Ade:2015lrj}.}\label{fig1}
\end{figure*}

\vspace{0.5cm}
{\it Observational restrictions:}
The 
point of this section is to test the viability of LQC 
at the inflationary level, involving the latest CMB data.
Concerning the number of e-folds we use $N=50$ and $N=60$ respectively.
Overall, we find that for the current parametrizations of the 
dissipation parameter 
our $(n_{s},r)$ predictions satisfy
the restrictions of Planck within $1\sigma$
uncertainties. For example, 
in figure 1 we present the $(n_{s},r)$
diagram which is provided by the Planck team. 
On top of that 
we plot the individual sets of $(n_{s},r)$
in LQC which is based on the dissipation parameter $\Gamma \propto T$.
Evidently, our results are consistent with those of Planck. 
Regarding the scalar spectral index we find  $n_{s}\simeq 0.962$ 
which is agreement with that of Planck 
$n_{s}=0.968\pm 0.006$. Lastly, as it
can be viewed from figure 1, the tensor-to-scalar fluctuation 
ratio could reach the value of $r\simeq 0.02$ 
which is consistent within $1\sigma$ with that of 
BICEP2/KeckArray/Planck results \cite{Keck15}.

To summarize, the inflationary class  
of LQC models seems to accomplish 
two main achievements: i)
it smoothly connects the Planck era with the CMB epoch;
and ii) it leads to the same successful 
prediction compatible with the CMB parameters 
$(n_{s},r)$
provided by the Planck collaboration 
\cite{Ade:2015lrj,Keck15}. 
Overall the combination of the recent work of Ashtekar and Gupt \cite{Gupt17} 
with the present Essay 
provides a complete investigation of LQC 
approach at the era of CMB.
From both works it becomes clear that LQC, which is the outcome of 
the effective LQG theory, could provide an efficient way to understand the 
early phase of the universe.

\end{document}